# Tuning Spin Transport in a Graphene Antiferromagnetic Insulator.

Petr Stepanov[1, 2*], Dmitry L. Shcherbakov[1], Shi Che[1], Marc W. Bockrath[1], Yafis Barlas[3], Dmitry Smirnov[4], Kenji Watanabe[5], Takashi Taniguchi[6], Roger K. Lake[7], Chun Ning Lau[1*]

[1] Department of Physics, The Ohio State University, Columbus, OH 43221.
[2] ICFO - Institut de Ciencies Fotoniques, The Barcelona Institute of Science and Technology, Castelldefels, Barcelona, 08860, Spain.
[3] Department of Physics, University of Nevada, Reno, Reno, NV 89557.
[4] National High Magnetic Field Laboratory, Tallahassee, FL 32310.
[5] Research Center for Functional Materials, National Institute for Materials Science, 1-1 Namiki, Tsukuba, 305-0044, Japan.
[6] International Center for Materials Nanoarchitectonics, National Institute for Materials Science, 1-1 Namiki, Tsukuba, 305-0044, Japan.
[7] Department of Electrical and Computer Engineering, University of California, Riverside, Riverside, CA 92521.

[*]Correspondence to: petr.stepanov@icfo.eu, lau232@osu.edu.

Long-distance spin transport through anti-ferromagnetic insulators (AFMIs) is a long-standing goal of spintronics research. Unlike conventional spintronics systems, monolayer graphene in quantum Hall regime (QH) offers an unprecedented tuneability of spin-polarization and charge carrier density in QH edge states. Here, using gate-controlled QH edges as spin-dependent injectors and detectors in an all-graphene electrical circuit, for the first time we demonstrate a selective tuning of ambipolar spin transport through graphene ν=0 AFMIs. By modulating polarities of the excitation bias, magnetic fields, and charge carriers that host opposite chiralities, we show that the difference between spin chemical potentials of adjacent edge channels in the spin-injector region is crucial in tuning spin-transport observed across graphene AFMI. We demonstrate that non-local response vanishes upon reversing directions of the co-propagating edge channels when the spin-filters in our devices are no longer selective for a particular spin-polarization. Our results establish a versatile set of methods to tune pure spin transport via an anti-ferromagnetic media and open a pathway to explore their applications for a broad field of antiferromagnetic spintronics research.

**Introduction**.

Achieving long distance (>>10nm) dissipationless spin-transport and understanding its microscopic mechanism are important goals of antiferromagnetic spintronics research [1–3]. A plethora of condensed matter systems has been recently proposed as active components of ferromagnetic spin-transport devices [4–8]. For example, in magnetically doped insulators spin-currents can be carried via magnon quasiparticles [9,10], while in easy-plane magnetically ordered systems, spin-currents can be carried via dissipationless mechanisms, such as spin-superfluidity [1,2,11–15]. Recently, monolayer graphene (MLG) in the quantum Hall regime (QH) has been realized as a highly tunable system for spintronic studies [16–19].

High magnetic field frequently breaks an approximate *SU(4)* symmetry of MLG, giving rise to a set of symmetry-broken Landau levels (LL) with integer fillings $\nu=\pm 2$, $\pm 1$ and 0 [20–33]. While states at fully filled or fully emptied zeroth Landau level ($\nu=+2$ and $\nu=-2$, respectively) host two co-propagating edge channels with opposite spin order, QH states at $\nu=\pm 1$ are spontaneously oppositely spin-polarized [21,24,26,34]. The half-filled ($\nu=0$) zeroth LL is a unique QH state and is now broadly recognized by theoretical and experimental research [20,21] that in the vicinity of high magnetic fields its ground state is a canted antiferromagnet (CAF), where charge quasiparticles occupy graphene's two sublattices with almost oppositely directed spins in an easy-plane configuration.

**Detecting spin-current using chiral QH edge states**.

Because of the small spin-orbit coupling and extremely small spin-anisotropy, QH states in graphene have been proposed as a platform for spintronics research [11,35]. In contrast to conventional spintronics systems [1,9], graphene offers a remarkable *in situ* tuneability of charge carrier density and spin-polarization in QH edge channels, when charge carriers' Fermi level can be capacitively tuned by changing electrostatic potential on the back gate electrode. By tuning the external fields, we demonstrate how chiral QH edge states switch the spin transport polarity. We utilize QH states within zeroth LL to detect a large non-local response $V_{nl}$. $V_{nl}$ is asymmetric upon reversing the bias polarity, which originates from different hot-spots generated by inter-channel scattering and may likely be explained by spin canting direction in CAF state. When the direction of the out-of-plane magnetic field is reversed, we observe a complete quenching of the non-local response, pointing towards a spin-filtering mechanism shut-off on both sides of the CAF state and explained by the reversed direction of the edge channel propagation. Lastly, we demonstrate the revival of the non-local signal using hole-like QH states in the same device.

Fig. 1a illustrates a schematics of QH states in a long (~15 μm) hBN-encapsulated graphene channel. On top of each sample we fabricate three metal top gates following the geometry proposed in ref. [11] and experimentally realized in previous works [16,17] (insets in Fig. 1a). Devices in this study were fabricated using a dry-transfer technique. First, we use a thin polypropylene carbonate (PPC) layer placed on top of the polydimethylsiloxane (PDMS) stamp to pick up a top hBN layer of a few tens of nanometers thick at 40 ºC. Next we pick up a graphene flake mechanically exfoliated on a $SiO_2/Si^{++}$ wafer, followed by the subsequent pick up of the bottom hBN flake, which fully encapsulates graphene between clean atomically flat hBN interfaces. At the last step, the stacks are placed on the target $SiO_2/Si^{++}$ substrate followed by the release of the PPC layer from the PDMS stamp at 90 ºC. The next step of fabrication consists of etching the stack into a multiple Hall-bar geometry using reactive plasma etching by a high-power $SF_6$ plasma. Next, we couple the graphene sheet to the Cr/Au metallic contacts (10/50 nm). At the last step, Cr/Au metallic top gates are deposited on top of the sample separated by a thin $Al_2O_3$ layer deposited beforehand to exclude shortages of the top gates with the graphene channel. In the end of the nanofabrication process, we obtain a graphene sample with fifteen leads for proper transport characterization, three top gates and one bottom gate to independently control charge carrier densities in four regions of the graphene channel.

We apply voltage to the Si back gate so the non-top-gated regions (blue regions in Fig. 1a) host QH states at fillings $\nu=\pm 2$. The central top gate tunes the Fermi level into the charge neutrality point with $\nu=0$. Left and right top gates serve as spin-filters in the spin-injection (red box

$\nu=+2/+1/+2$) and spin-detection (blue box $\nu=+2/+1/+2$) regions, respectively. Spin selectivity is defined by the QH state in the side-top-gated regions. A bias voltage is applied between the inner and outer $\nu=2$ sections in the injection region, while a non-local voltage $V_{nl}$ is measured between the inner and outer $\nu=2$ sections in the detection region.

In this experiment, edge state configurations play a crucial role in controlling the spin current injection/detection processes. Fig. 1b shows QH states inside the spin-injector and spin-detector regions. When bias is applied between the $\nu=2$ sections in the injection region, the high-energy ("hot") carriers propagate from source to drain electrode (marked **S** and **D**, respectively); at the corner of the spin-filtering $\nu=+1$ QH state (pink circle), only the ↑-spin polarized (red) channel of the outer $\nu=+2$ QH state is permitted into the $\nu=+1$ QH state region without loss of momenta, thus undergoing the spin-filtering process. The yellow circle indicates another pivot point, where the previously selected ↑-spin polarized (red) channel enters the inner $\nu=+2$ QH state further generating a spin angular momentum imbalance between adjacent co-propagating edge channels. The ↓-spin polarized (blue) channel of the inner $\nu=+2$ QH state carries only "cold" carriers as defined by the edges of the spin filtering $\nu=+1$ QH state and the direction of the *B*-field. On the right-hand side of the device (blue box in Fig. 1a), spin detection is achieved via a reciprocal process. Here the voltmeter probes **A** and **B** are connected to the inner and outer $\nu=+2$ regions to detect a non-local signal $V_{nl}$. Since the central $\nu=0$ region does not carry a charge current, a finite $V_{nl}$ indicates the presence of pure spin current through the graphene AFMI [16]. Specifically, it directly probes the chemical potential imbalances ($V_A–V_B$), and the sign of the detected signal indicates a dominant spin polarization: the voltage probe **A** detects an excess of the ↑-spin and **B** detects an excess of the ↓-spin angular momentum. Since in this experiment we also change orientation of the magnetic field, we choose an assignment of the spin directions shown in Fig. 1b as a reference for the rest of the data shown in this study, i.e. we define spin-up and spin-down as out of page and into page, respectively. We define the magnetic field and charge carrier spin directions with respect to Fig. 1b, assuming that electron-like states misalign their spin with the magnetic field thus occupying a higher energy broken symmetry state, while hole-like QH states align their spin with the magnetic field thus acquiring an opposite to electrons spin-polarization [21].

**Magnetoresistance data**.

An essential ingredient for such spin-signal detection is a development of QH plateaus in transverse magneto resistance data $R_{xy}$, which indicate a formation of topological edge currents in the sample edges. For this purpose, we perform the $R_{xx}$ and $R_{xy}$ measurements shown in Fig. 2. Fig. 2a and 2b demonstrate contour plots that clearly exhibit Landau fan diagrams for longitudinal and transverse resistances, respectively. Fig. 2c and 2d show the linecuts taken at $B=18$ T and $-18$ T. Here, we observe well-developed QH plateaus with $R_{xy}$ taking discrete values at $h/e^2$, $h/2e^2$, $h/3e^2$…, where $h$ is Plank´s constant and $e$ is elementary change, while $R_{xx}$ vanishes, overall signaling QH edge channel formations.

We precisely identify positions of different QH plateaus on the $V_{TG}$-$V_{BG}$ map. For this purpose we measure $R_{xy}$ as a function of top gate and back gate voltages at constant $B=18$ T (Fig. 3). A set of stripes of the same quantization numbers emerge for finite ranges of the gate voltages. The stripes that do not exhibit dependence on the top gate voltage (horizontal dashed lines) correspond to a single-gate-like geometry, where only the back gate remains effective. These back gate values are used to tune the Fermi level within the ¨bare¨ regions of the sample. The stripes that exhibit dependence on the top gate voltage, in contrast, correspond to the regions where both gates are effective and these values are utilized to tune the Fermi level in the top-gated regions of the sample.

**Bias dependence of spin-transport signal.**

Electrons residing on graphene sublattices **A** and **B** in the CAF state acquire oppositely directed spins with a small out-of-plane canting caused by the Zeeman effect (Fig. 1c) [21]. When impinging upon the CAF, the incident spin-current on the left side of the CAF favors the formation of the spiral Neél texture that carries spin current collectively [11]. During the measurements, the Fermi levels of the bare ($\nu=+2$) and central top gated ($\nu=0$) regions are fixed, and the filling factors under the side top gates inside the injection and detection regions ($\nu_{inj}$ and $\nu_{det}$, respectively) are varied. Typical data is shown in Fig. 1d, which demonstrates $V_{nl}$ vs. $\nu_{inj}$ and $\nu_{det}$ at $B=18$ T and $T=1.8$ K. We observe a large non-local response $V_{nl} \sim 170$ μV (see inset $V_{nl}$ vs. $\nu_{det}$), only when both the injection and the detection top-gated regions' Fermi levels are set at $\nu=+1$ QH states. The details of this experiment are given elsewhere [16].

We now turn our attention to tuning the observed non-local signal via available experimental knobs. Since the spin angular momentum imbalance between ↑- and ↓-spin-polarized edge channels is vital for spin current propagation, we choose to explore $V_{nl}$ vs. $\nu_{det}$ and $\nu_{inj}$ under a reversed S-D bias voltage. Fig. 4a shows non-local measurements at $\mu < -E_z$ ($V_{bias}=-65$ mV). We observe a clear suppression of the non-local signal. The non-local signal demonstrates a weakly negative response $V_{nl} \sim -30$ μV when both $\nu_{det}=\nu_{inj}=+1$ (see inset). We further observe the asymmetry in $V_{nl}$ by varying the **S**-**D** bias in Fig. 4c. Here we fix $\nu_{inj}=+1$ and measure $V_{nl}$ as a function of $\nu_{det}$ and $V_{bias}$. Our data demonstrates a shift of maximized non-local voltage for the negative $V_{bias}$ compared to the positive $V_{bias}$ (yellow stripes in Fig. 4d). The maximum signal ($V_{nl}$) under positive bias occurs at $V_{bias}=65$ mV ($\mu > E_z$) with magnitude $V_{nl} \sim 170$ μV, while the maximum signal under negative bias occurs at a substantially larger $-V_{bias}=240$ mV ($\mu < -E_z$), with a smaller magnitude signal, $V_{nl} \sim 110$ μV. Suppression of the non-local signal is inconsistent with the scenario when the inter-channel scattering between the adjacent QH edge states is neglected.

We explain these observations by examining the edge state configuration shown in Fig. 4b. At negative bias, the hot edge that carries more spin-polarized charge current is propagating from **D** to **S** and only residual QH current propagates further through the ↑-spin channel. The hot ↑-spin-polarized QH edge state does not undergo a spin-filtering process through the $\nu_{inj}=+1$ QH state region as it equilibrates on the source electrode, and the spin angular momentum imbalance on the edge of CAF is dominated by the hot ↓-spin channel inside the inner $\nu=+2$ in the spin-injector region. We note that the biases applied to our devices are larger than the Zeeman energy splitting between the ↑- and ↓-spin-polarized edge channels ($|\mu|=|eV_{bias}|>E_z \sim 2.08$ meV at 18 T). This implies that the electrons in the hot edge acquire enough energy to overcome the energetic barrier defined by the spatial displacement of the edge channels´ wave functions [36,37] and flip their spins at localized "hotspots" (light-blue circles in Fig. 1b and Fig. 4b, respectively). This allows for an extra angular momentum to propagate inside the bulk of $\nu=+1$ QH state [10,11,38–42]. We note that only ↓-spin polarized angular momentum can propagate into a ↑-spin polarized bulk. Furthermore, an extra angular momentum can be absorbed by the QH edges on the opposite corners of $\nu=+1$ QH state resulting in a spin-flip via a reverse process. It has been previously demonstrated that such spin transfer is mediated by magnons [17]. Therefore, the location at which the magnons are launched due to the inter-channel scattering process becomes important. For positive $V_{bias}$ (Fig. 1b) it happens farther from the edge of the CAF state, meaning the effect of the inter-channel scattering is insignificant. In the case of negative $V_{bias}$ (Fig. 4b), the loss of spin angular momentum happens closer to the injection site thus suggesting a larger loss of the momenta at the edge of the CAF and leading to a smaller non-local response.

Edge channel equilibration plays an important role in the studied devices. We would like to highlight that $n$-$n´$-$n$ ($p$-$p´$-$p$) junctions in the injector and the detector regions have lateral dimensions smaller (~1 μm) than expected for partial or full equilibration between graphene QH state wave functions [43]. Therefore, the spin-channels in the inner and the outer $\nu=+2$ regions are not equilibrated and experience a spin-flip process assisted by the angular precession of the ferromagnetic bulk only at the designated spots (marked light-blue circles). In addition to Gilbert damping in AFMIs [12], the described mechanisms of angular momentum loss have a dramatic effect on the efficiency of the spin-transport. Taken the ratio $V_{nl}/V_{bias}\sim 10^{-3}$, our approximate estimation yields [16,44] a spin-transport efficiency parameter $\sim 10^{-2}$-$10^{-4}$, which is in a good agreement with previous studies in AFMIs [45–47].

**Switching-off the spin-current in graphene AFMIs**.
Next, we manipulate the spin-current by utilizing a $B$-field of the opposite direction. Fig. 5a shows $V_{nl}$ vs. $\nu_{det}$ and $\nu_{inj}$ at reversed magnetic field $B=-18$ T at $T=0.5$ K ($V_{bias}=65$ mV). Interestingly, the non-local signal is completely quenched $V_{nl}\sim 0$ μV indicating that no spin-current propagates through AFMI (Fig. 5a). We attribute this observation to the equilibration of spin angular momentum in the ↑-spin- and ↓-spin-polarized edge channels in the inner $\nu=+2$ on the left edge of the CAF. Under reversal of the magnetic field, the QH edge channels change their direction of propagation from counter-clockwise (Fig. 1 and Fig. 4) to clockwise (Fig. 5b). In this case, both co-propagating inner $\nu=+2$ "cold" edge channels originate from the drain electrode avoiding spin-selective $\nu=+1$ QH state. This results in the absence of net spin on the edge of $\nu=0$ AFMI and, therefore, absence of non-local response in the detection region of the device.

A similar effect of spin current shut-off can be achieved by switching sides of the detector and the injector regions in the original configuration of Fig. 1 and 4 (see Fig. 6). Just like in the case of Fig. 5, we observe a clear quenching of spin transport. Similarly to the case of reversed magnetic field polarity, spin angular momentum is absent on the edge of CAF as the "cold" charge carriers do not undergo spin-filtering process. They emanate from the drain electrode maintaining no spin angular momentum imbalance on the edge of CAF state (now on the right-hand side of CAF state).

**Utilizing hole-like QH states to manipulate spin-tranport**.
Finally, taking advantage of the ambipolar charge transport in graphene, we examine spin-transport using hole-like QH states. Fig. 7a demonstrates an example of a measurement performed using a hole-like set of QH states at LLs with fillings $\nu=-2/-1/-2/0/-2/-1/-2$. The spin filtering and transport mechanisms in the case of hole-like states are identical to the former case of electron-like states. Similar to $\nu=+1$, the QH state at $\nu=-1$ is a ferromagnetic insulator with one spin polarized edge channel. In order to detect spin-current signals, we have to account for the sign of quasiparticle charges in addition to the magnetic field and bias directions. A schematic image shown in Fig. 7b illustrates the configuration of the edge states in the injection region for the experiment shown in Fig. 7a. Hole-like quasiparticles move counter-clockwise undergoing spin-selection through a $\nu=-1$ QH state resulting in the spin chemical potential imbalance at the edge of the graphene AFMI similar to Fig. 1c. In addition, we change the bias sign $V_{bias}=-65$mV ($\mu < -E_z$) to select the ↓-spin-polarized edge channel. As a result, we observe a strong non-local signal $V_{nl}\sim-100$ μV (Fig. 4a) when $\nu_{det}=\nu_{inj}=-1$. In the case of a reversed magnetic field polarity for the hole-like QH states (Fig. 8), we expectedly observe a disappearance of the non-local signal owning to the spin-angular momentum absence on the edge of the CAF state. This scenario is very similar to the case of the electron-like QH states shown in Fig. 5, where the spin-filtering processes are shut down and the edge channels at the left-hand side of the graphene AFMI originate from the **D** electrode.

**Conclusions**.

To conclude, our results demonstrate a high sensitivity of the pure spin transport through a $\nu=0$ graphene AFMI to various experimental parameters. The "on" and "off" states of the non-local response are dependent on the edge state propagation direction that can be tuned by a number of direction-sensitive parameters, such as magnetic field orientation, polarity of biasing voltage, and chirality of the charge quasiparticles. These observations demonstrate independent manipulation of spin current that may be carried by either electrons or holes, and further expands graphene as a versatile platform to study antiferromagnetic and ferromagnetic spintronics. Further experiments would be required to establish length, width, mobility dependencies of non-local signal responses. A number of additional experiments will be required to achieve a nearly dissipationless spin-transport in practical magnetic materials for information processing and storage applications.

**Sign convention.** We choose to assign the spins in the spin-resolved edge channels as following: we opt in to defining ↑- (↓-) spin direction as out of (into the) page. Magnetic field directions are chosen in an identical way: up (down) *B*-field is chosen out of (into the) page. We choose the excitation bias values to match chemical potential signs for both electrons and holes.

**Acknowledgements** We thank Yizhou Liu for useful discussions. The experiments are supported by DOE BES Division under grant no. DE- SC0020187. Device fabrication is partially supported by the Center for Emergent Materials: an NSF MRSEC under award number DMR-1420451. A portion of this work was performed at the National High Magnetic Field Laboratory, which is supported by National Science Foundation Cooperative Agreement No. DMR-1157490 and the State of Florida. Growth of hexagonal boron nitride crystals was supported by the Elemental Strategy Initiative conducted by the MEXT, Japan (Grant Number JPMXP0112101001) and JSPS KAKENHI (Grant Numbers JP19H05790 and JP20H00354).. P.S. acknowledges support from the European Union's Horizon 2020 research and innovation programme under the Marie Skłodowska-Curie grant agreement No. 754510.

**Correspondence and requests for materials** should be addressed to C. N. L, and P. S.

**Competing interest.** Authors claim no competing interest.

**Figures.**

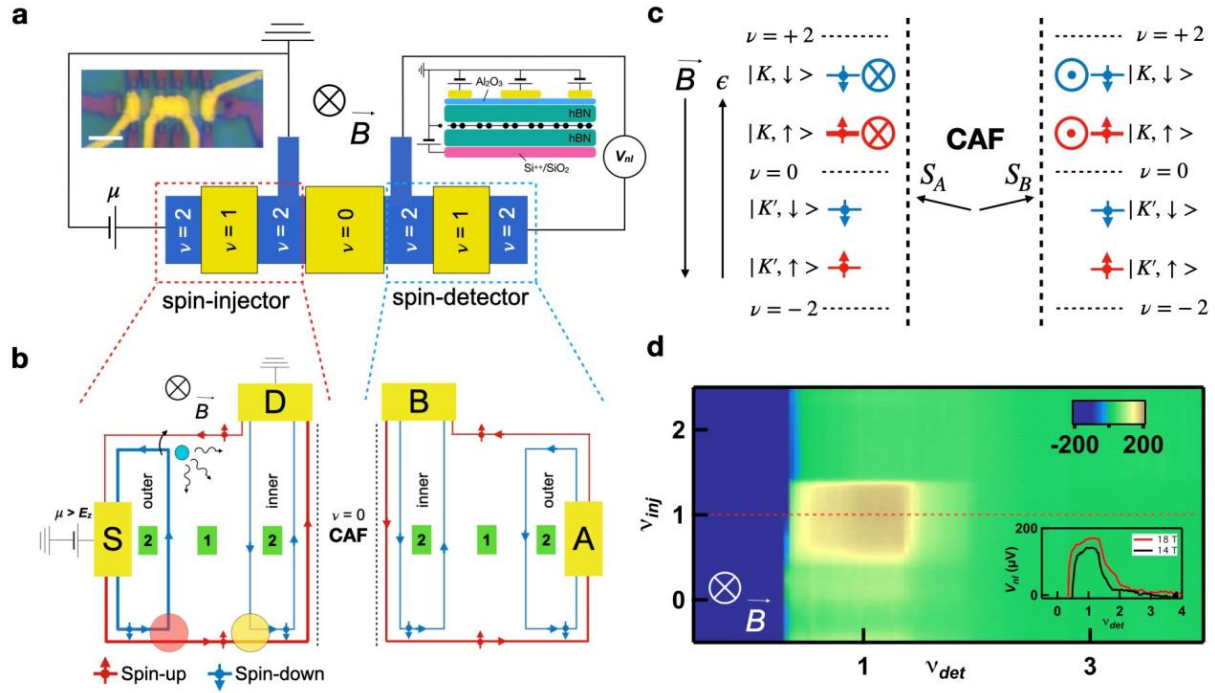

Figure 1. | **Spin transport detection. a**) All-graphene electric circuit schematics. Blue regions are "bare" back-gated. Yellow regions are gated independently with both bottom and top gates. Dashed red- (blue-) line boxes show injection (detection) region, respectively. The left inset demonstrates an optical micrograph of a typical device (scale bar is 5 μm). The right inset shows the heterostructure profile that consists of an encapsulated MLG layer and three metal top gates. **b**) QH edge states inside the spin-injection and spin-detection regions. Solid red (blue) lines demonstrate QH state edge channels with ↑(↓)-spin polarization. Thicker (thinner) lines refer to the "hot" ("cold") edges, that carry high (low) chemical potential, respectively. The direction of the magnetic field is chosen to be into the plane of the figure leading to the counter-clockwise propagating electron-like edge channels. **c**) Detailed configuration of zeroth LL states near graphene easy-plane AFMI ν=0 region. CAF state is sandwiched between the two inner QH states at fillings ν=+2 that host two co-propagating edge channels (marked by circles) with spin angular momentum imbalance indicated by a thicker ↑-polarized red line. **d**) Experimental measurements of the non-local signal (in μV) at $B$=18 T and $T$=1.8 K. The inset in the right lower corner shows line traces at $B$=18 T (red) and $B$=14 T (black) taken along the red dashed line.

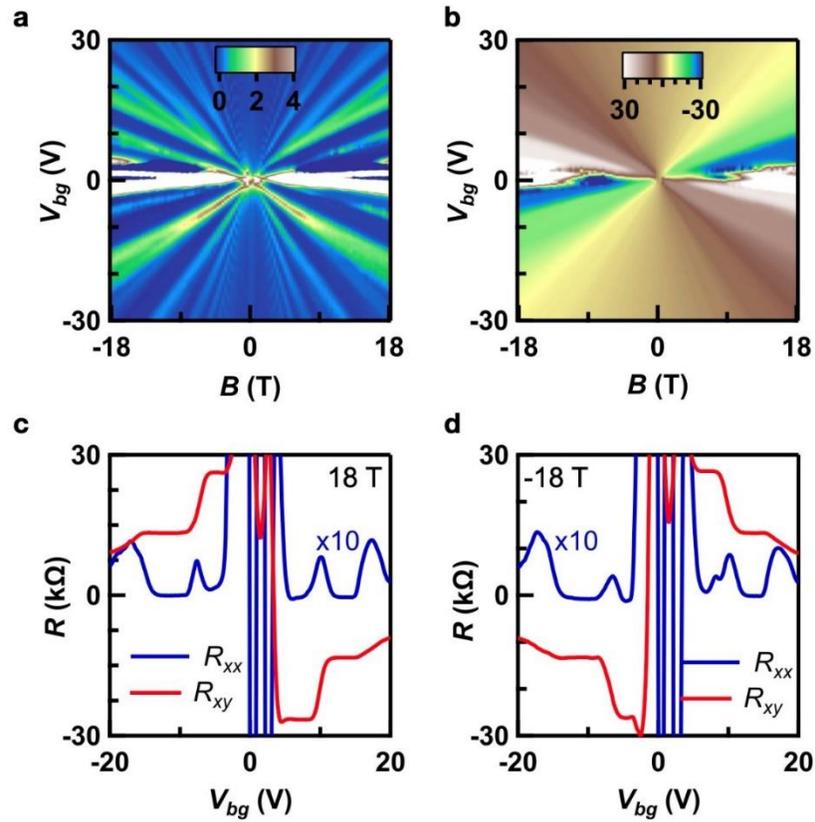

Figure 2. | **Magnetoresistance measurements and QH effect in the graphene monolayer device. a)-b)** Contour plots of $R_{xx}$ and $R_{xy}$ (in kΩ) measured as a function of back gate voltage and magnetic field ($V_{tg}$=0 V) in the central region of the device using a Hall-bar geometry. **c)-d)** $R_{xx}$ and $R_{xy}$ vs. $V_{bg}$ taken at 18 T and -18 T respectively. Linecuts demonstrate high quality QH effect with well-developed plateaus within the zeroth LL of graphene. $T$=0.5 K.

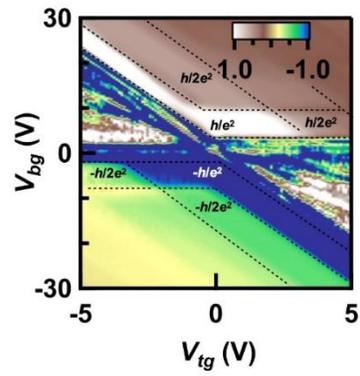

Figure 3. | **Magnetoresistance measurements of $R_{xy}$ (in units of $h/e^2$) at constant $B$=18 T.** This data is used to identify proper positions of QH plateaus on the gate voltage map, which is further used to tune the "bare" and top-gated regions into QH states of interest.

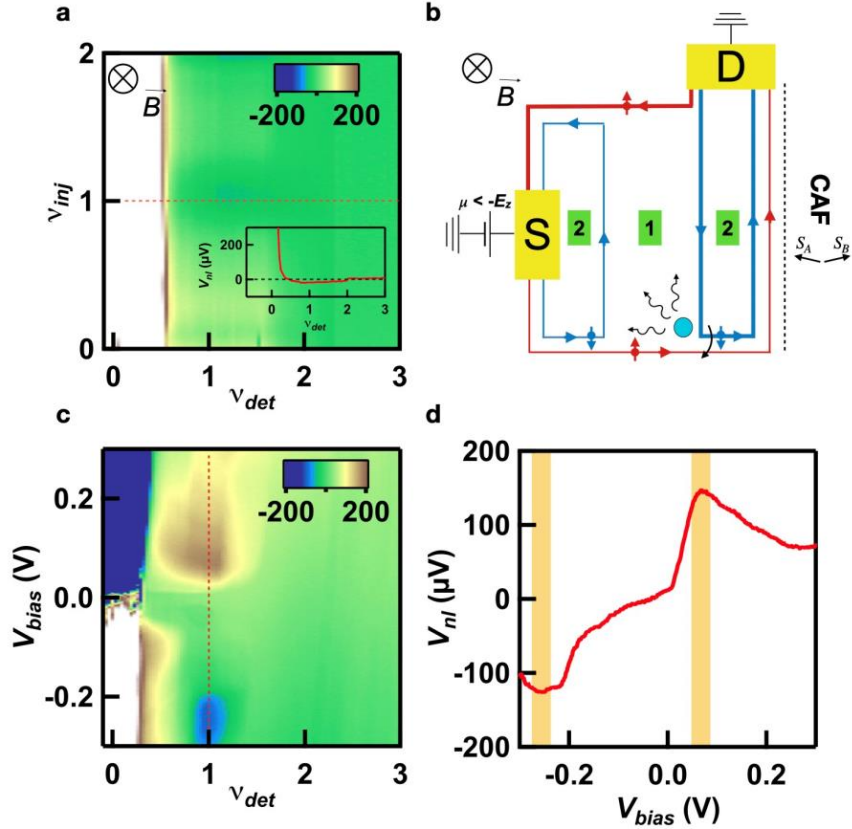

Figure 4. | **Demonstration of spin-current switch using a reversed bias circuit. a)** $V_{nl}$ (in μV) under reversed bias $V_{bias}$=-65 mV ($\mu$<-$E_z$) at $T$=0.5 K. Inset shows a line cut taken along the red dashed line. Note a significantly suppressed non-local signal around $\nu_{det}$=+1. **b)** Schematic image of QH states configuration under the inverted voltage bias. Note the change in "hot" and "cold" edge state configuration compared to Fig. 1b. The light-blue circle indicates magnons launched into the ferromagnetic bulk. **c)** $V_{nl}$ (in μV) as a function of $\nu_{det}$ and $V_{bias}$- at a fixed $\nu_{inj}$=+1. **d)** $V_{nl}$ vs. $V_{bias}$ taken along the $\nu_{det}$=+1 in c). Yellow stripes indicate maximized $V_{nl}$. We observe a significant asymmetry in the acquired data.

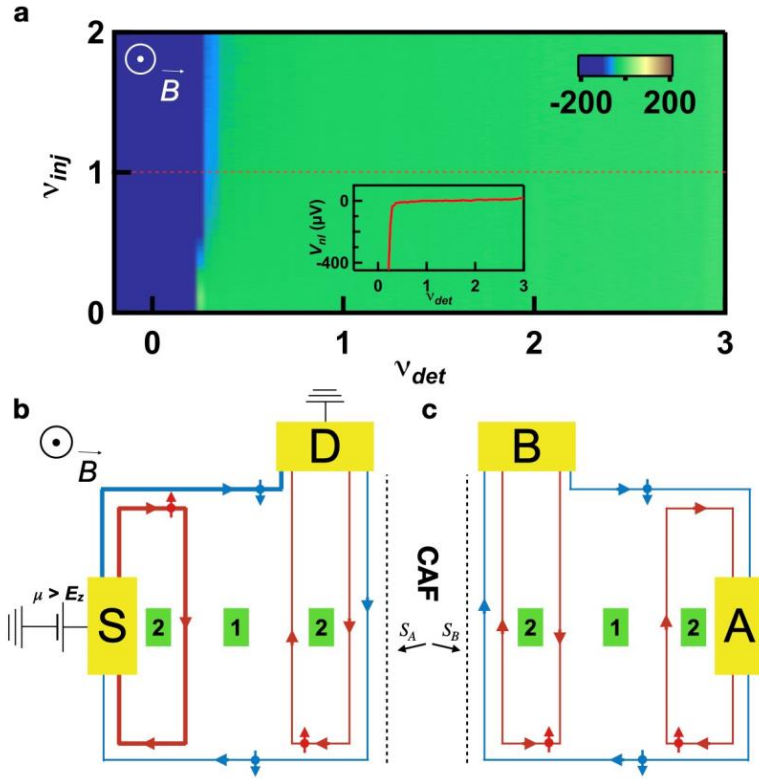

Figure 5. | **Demonstration of the spin-current switch using an oppositely directed magnetic field. a)** $V_{nl}$ (in µV) under the inverted $B=-18$ T at $T=0.5$ K. Inset shows a line cut taken along the red dashed line. **b)-c)** Schematic images that explain QH edge state configuration under the inverted magnetic fields inside the spin-injection (b) and spin-detection (c) regions. Non-local signal quenching is attributed to the absence of the spin angular momentum on the edge of the CAF state since both co-propagating channels in the inner ν=+2 region originate from the same ¨cold¨ electrode **D**.

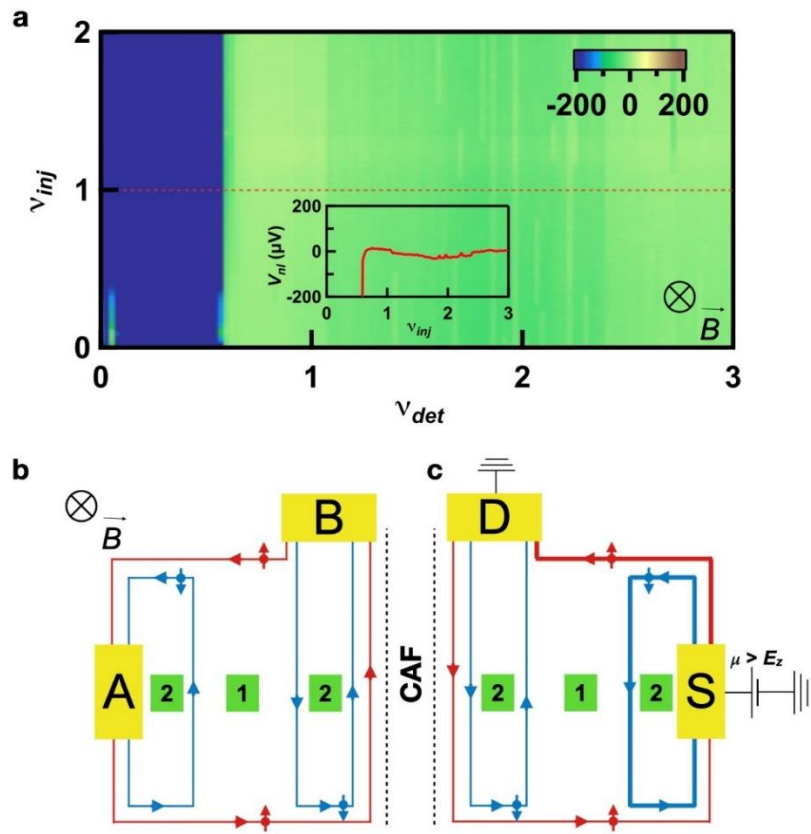

Figure 6. | **Effect of the reversed spin injector/detector sides. a)** Measurements of the non-local response under inverted spin injector/detector sides of the device in contrast to shown in Fig. 1b. $B$=18 T and $T$=0.5 K. Inset shows a line trace taken along the red dashed line. **b)-c)** Schematic images of the QH state configuration inside the spin-detector (b) and spin-injector (c) regions.

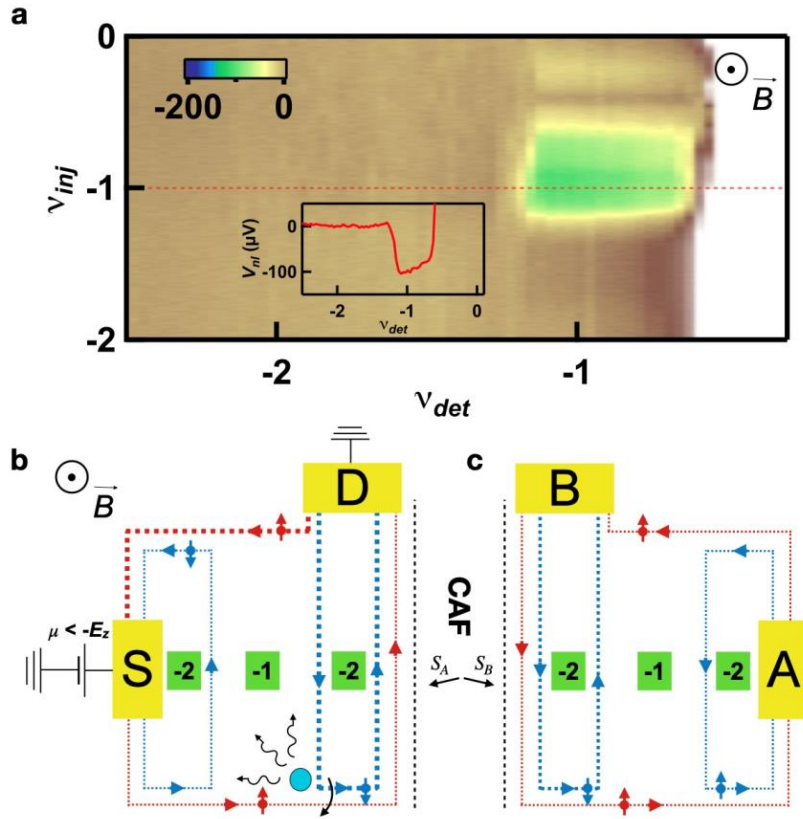

Figure 7. | **Utilizing hole-like LLs for spin-current detection.** **a)** Hole-side non-local measurements (in μV) demonstrate strong spin current response around $\nu_{det}=\nu_{inj}=-1$. Inset shows a line trace taken along the red dashed line. **b)** Schematic images of the edge states configuration in the injection and **c)** the detection regions in the case of measurements shown in a). Note that here we detect the oppositely directed spin-current compared to Fig. 1d due to the higher chemical potential of the ↓-spin channel in the inner $\nu=-2$ region. Dotted red and blue lines indicate hole-like QH edge states with ↑- and ↓-spin polarizations, respectively.

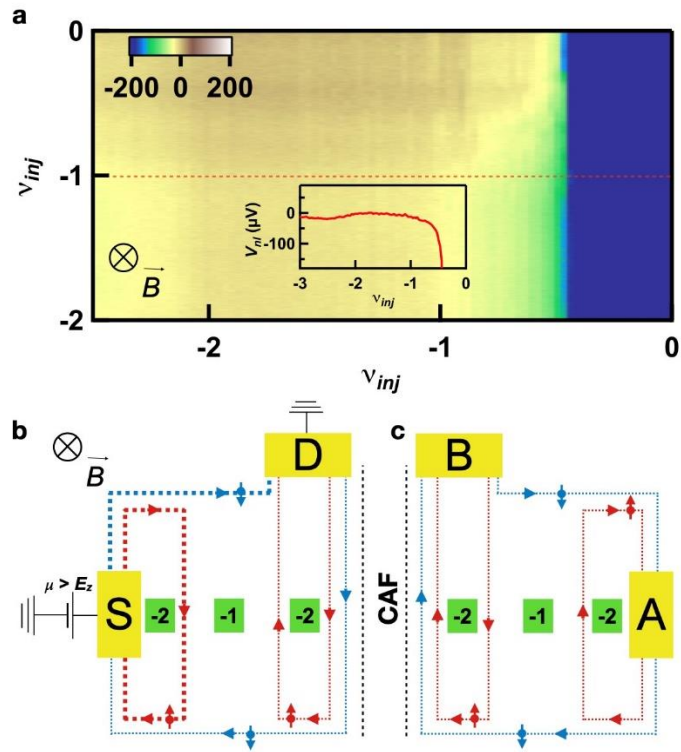

Figure 8. | **Non-local measurements for the hole type carriers at *B*=18 T. a)** Measurements of the non-local response for the hole type carriers in Fig. 1d. Inset shows a line trace taken along the red dashed line. **b)-c)** Schematic images of the QH state configuration inside the spin-injector (b) and spin-detector (c) regions. The absence of the non-local response is explained by the reversed current direction in the edge channels of the QH states for the hole type carriers compared to electron-like (Fig. 1d), which leads to the absence of spin angular momentum on the left edge of CAF in the injector region side (b).